\documentclass{emulateapj}
\usepackage{apjfonts}
\def\msun{{\rm ~M}_{\odot}}

\def\mpy{{\rm ~M}_{\odot} {\rm ~yr}^{-1}}

\def\racc{{R_{\rm acc}}}
\def\rcap{{R_{\rm cap}}}
\def\vrel{{v_{\rm rel}}}
\def\cs{{c_{\rm s}}}
\def\nasr#1{{\sl New \ Astr. Rev.,} { #1}}
\def\lapprox{\;\rlap{\lower 2.5pt                       
             \hbox{$\sim$}}\raise 1.5pt\hbox{$<$}\;}
\begin{document}

\title{The Interaction of Stellar Objects within a Common Envelope} 

\author{Paul M. Ricker\altaffilmark{1}
and Ronald E. Taam\altaffilmark{2}}

\altaffiltext{1}{Department of Astronomy, University of Illinois, 1002 West Green Street, Urbana, IL 61801; pmricker@uiuc.edu.}
\altaffiltext{2}{Department of Physics and Astronomy, Northwestern University, 2131 Tech Drive, Evanston, IL 60208; r-taam@northwestern.edu.}

\begin{abstract}
We use high-resolution, three-dimensional hydrodynamic simulations to study
the hydrodynamic and gravitational interaction between stellar companions
embedded within a differentially rotating common envelope.  We evaluate the
contributions of the nonaxisymmetric gravitational tides and ram pressure forces
to the drag force and, hence, to the dissipation rate and the mass accumulated
onto the stellar companion.  We find that the gravitational drag dominates the
hydrodynamic drag during the inspiral phase, implying that a
simple prescription based on a gravitational capture radius
significantly underestimates the dissipation rate and overestimates the inspiral
decay timescale.  Although the mass accretion rate fluctuates
significantly, we observe a secular trend leading to an effective rate
that is significantly less than the rate based on a gravitational capture
radius. We discuss the implications of these results within the context of
accretion by compact objects in the common-envelope phase. 
\end{abstract}


\section{Introduction}
To understand the evolution of binary star systems, it is essential to analyze 
the interactions between their stellar components. Examples of 
such influences include the spin-orbit tidal interaction and mass transfer,
as well as interactions that result in the loss of mass and angular momentum.
Equally important are the interactions of stars orbiting about
their common center of mass within a differentially rotating common envelope.
It is generally accepted that such an evolutionary stage is essential for the
formation of short-period binary systems containing compact objects (see, e.g.,
Iben \& Livio 1993 and Taam \& Sandquist 2000). In this case, the interaction
determines the orbital evolution of the system and the conditions under which
the common envelope is ejected, leading to the survival of a remnant binary system
or to a merger that forms a rapidly rotating single star.  The amount of mass and
angular momentum accreted by the inspiralling components during this
phase also has direct implications for the properties of
the compact object population in binary systems. 

Lacking multidimensional hydrodynamical simulations of 
the common-envelope phase, the initial numerical and semi-analytical 
studies of the problem used simple prescriptions for the stellar interactions 
based on the pioneering work by Hoyle \& Lyttleton (1939) and Bondi \& Hoyle (1944), as 
generalized by Bondi (1952).  These seminal studies focused on the idealized 
problem of the capture of matter by a gravitating point object moving supersonically 
with respect to a uniform medium.  In this framework, a gravitational capture radius, 
$\rcap$, plays an important role in determining the rates of mass accretion and 
energy dissipation.  $\rcap$ is given by
\begin{equation}
\label{Eqn:rcap} \rcap = {2 G M \over \vrel^2 + \cs^2}\ ,
\end{equation}
where $M$ is the mass of the gravitating object, $\vrel$ is the velocity of the 
object with respect to the medium, and $\cs$ is 
the local speed of sound.  When a density gradient with scale height $H$ is
present, the effective accretion radius $\racc$ is (Dodd \& McCrea 1952)
\begin{equation}
\label{Eqn:racc} \racc = {\rcap \over 1 + (\rcap/2H)^2}\ .
\end{equation}
The energy dissipation rate is then $L_{\rm d} \approx \pi
\racc^2 \rho \vrel^3$, where $\rho$ is the upstream density.  To improve on
these estimates,
hydrodynamic effects were approximated analytically by Ruderman \& Spiegel
(1971), Wolfson (1977), and Bisnovatyi-Kogan et al.\ (1979) as well as
numerically by Hunt (1971, 1979), Shara \& Shaviv (1980), and
Shima et al.\ (1985).  These early multidimensional simulations considered
axisymmetric flow, and their results have been used to calibrate the energy
loss rate.  In particular, the drag coefficients obtained from such simulations
(see, e.g., Shima et al.\ 1985) have been used to estimate the rate of energy
dissipation in the common envelope. 

However, many of the simplifying assumptions underlying these studies are
inadequate for direct application to common-envelope interactions.
The flow is nonaxisymmetric and distinctly nonuniform, reflecting the
existence of velocity or density gradients (the density may span several scale
heights within $\racc$). The effect of relaxing these assumptions
has been studied in two dimensions by Fryxell \& Taam (1988) and Taam \&
Fryxell (1989) and in three dimensions by Sawada et al.\ (1989) and Ruffert
(1999).  These studies could not encompass the full complexity of
common-envelope interactions, since the envelope's self-gravity was ignored.
Furthermore, because the companions move in elliptical
orbits, their cores interact with matter that has already been affected
in previous orbital phases. Thus, the state of the gas and its environment
in these calculations must be regarded as highly idealized.

Within the past decade, three-dimensional numerical studies of the common-envelope
phase that have relaxed earlier geometrical assumptions have been carried
out by Sandquist et al.\ (1998, 2000), DeMarco et al.\ (2003a,b), and Taam \&
Ricker (2006).  Recently, we have carried out high-resolution adaptive mesh
refinement (AMR) simulations of common-envelope evolution with effective
resolutions of $2048^3$ (R. E. Taam \& P. M. Ricker, in preparation), allowing the interaction of the
stars within the common envelope to be examined and quantified. 

In this Letter we report on some results of our numerical studies.  We focus on
analyzing a single high-resolution simulation to determine the hydrodynamic and
gravitational contributions to the drag forces affecting the orbital motion of
the stellar components during the early inspiral phase.  We also analyze the
accumulation of mass by the stellar components within the common envelope to
compare their magnitudes to estimates based on an accretion radius
formalism.  In \S~2, we briefly describe the numerical method and our assumed
model for the binary system.  Descriptions of the method of analysis and the
numerical results are presented in \S~3.  Finally, we summarize our results and
comment on their possible implications for applications involving compact
objects in short-period binary systems. 

\section{Numerical Method}
For our simulations we use FLASH, a parallel AMR code that 
includes both grid- and particle-based numerical methods (Fryxell et al.\ 2000).  
The Euler equations describing the evolution of the common-envelope gas are solved 
using the piecewise parabolic method (Colella \& Woodward 1984).  We use AMR 
to refine the mesh on the basis of the second derivative of gas density, for shock capturing, 
and on the basis of the positions of the stellar components, for high force resolution in their 
vicinities. This method significantly improves on stationary nested-grid techniques 
because the refined region can change to track the motion of both stars, thus
allowing investigations of systems with components of similar mass.

However, even with AMR the core of an evolved red giant star (the progenitor 
of a white dwarf) is too compact compared to the size of the computational domain 
(a factor of 40,000 smaller) given the large number of timesteps we require.  Since 
the core has a much higher mean density than does the common envelope, its strongest 
interaction with the gas is gravitational.  Therefore, we model the red giant core 
and the companion object using spherical particle clouds containing $2\times10^5$ 
particles each, obtaining forces on the clouds using cloud-in-cell (CIC) interpolation. 
(We use clouds rather than single particles to avoid problems related to CIC force 
anisotropy.) The cloud radii (ie., $R \sim R_\odot$) are taken to be 3 times the 
smallest zone spacing.  The particles in each cloud move rigidly together with the 
cloud's center of mass. More details are given by Taam 
\& Ricker (2006).

We considered the common-envelope evolution of a binary system consisting of a $1.05 
\msun$ red giant having a $0.36 \msun$ core and a $0.6 \msun$ companion with an orbital 
period of 44.2 days. The initial red giant model was taken from a one-dimensional stellar 
evolution code developed by Eggleton (1971, 1972).  We interpolated this model onto a 
three-dimensional Cartesian grid with nine levels of refinement. Since each block contained $8^3$ zones, and since the coarsest level contained one block, this corresponded to a minimum zone 
spacing of $2\times10^{10}$~cm (a factor of 2048 smaller than the domain size).  We 
artificially damped transient motions in the red giant envelope, evolving this model for 
a dynamical timescale. We then turned off the damping and added the particle cloud representing 
the companion star. At this point, the red giant core and envelope, as well as the companion,
were given circular orbital velocities and a spin angular velocity equal to 95\% of 
the synchronous value. 

\section{Numerical Results}
After an initial phase of 12~days, the orbital separation decreased monotonically 
with time, with an increasing rate of orbital shrinkage.  During the later phase 
($\sim$ 43~days), the orbital separation decreased by a factor of 3 from $4.3 \times 
10^{12}$ to $1.4 \times 10^{12}$~cm, as a result of the strong drag forces
on the cores.  In the physical system, the drag has both hydrodynamic and 
gravitational components, and, in our simulation, the gravitational component dominates. 

This can be seen by defining a measured hydrodynamic drag force $F_{\rm d}$ 
on each core via
\begin{equation}
\label{Eqn:flux}{\bf F_{\rm d}} = \int \rho {\bf v_{rel}} ({\bf v_{rel}} \cdot {\bf n}) dA\ ,
\end{equation}
where the integration is taken over a closed spherical surface centered on the core and 
{\bf n} is the unit vector perpendicular to its surface.  While the particle clouds 
have definite radii, these are determined by numerical constraints and do not correspond to 
the physical radii of the objects that the clouds represent. Moreover, the physical radii 
can be much smaller than the finest
zone spacing, depending on the type of companion star. Hence, we studied the effect of 
choosing different (resolvable) radii for the spherical surfaces used to measure 
the drag, ranging from $3.5\times10^{10}$ to $1.0 \times10^{12}$~cm.
We found that the hydrodynamic drag on the companion reaches a characteristic 
early inspiral value (at 34 days) of $\sim 9\times 10^{32}$~dynes at a radial scale
of $3\times 10^{11}$~cm from its center, only $\sim 10$\% of the gravitational drag
on this object. At this radial scale, the hydrodynamic drag on the red giant core
is at most $4\times10^{32}$~dynes, about 1/4 of the gravitational drag on it.
In general, the total
gravitational drag is one to two orders of magnitude larger than the hydrodynamical 
drag.  The drag force causes the orbital energy to dissipate at a rate given by 
the line integral of the drag force along the orbital path; in the case presented here,
the gravitational drag corresponded to a dissipation rate $L_{\rm d} \sim 8 \times 
10^{39}$ ergs s$^{-1}$.

\begin{figure}
\plotone{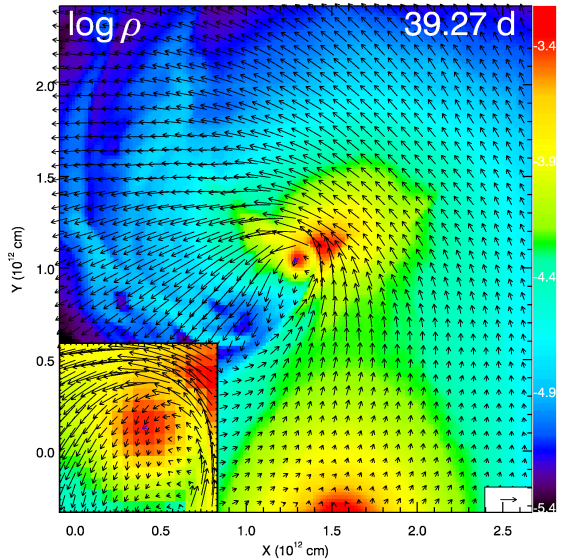}
\caption{\label{Fig:flow}
Flow field near the companion
at $39.27$~days.
Colors denote the logarithm of gas density in the $x-y$ plane, in units of g~cm$^{-3}$.
Arrows indicate the gas velocity in every fourth zone. The arrow in the lower
right-hand corner corresponds to 106~km~s$^{-1}$. The blue arrow at the center
indicates the companion's velocity. The red giant core lies just
off the bottom of the plot. The inset shows the region immediately surrounding the
companion, enlarged 3$\times$ per dimension. The vector scale is the same as in
the main image.}
\end{figure}

The flow near the inspiraling companion leads to the accretion of both mass 
and angular momentum. The flow pattern and density distribution in this region are
illustrated in Figure~\ref{Fig:flow}.  The density increases as one approaches the embedded object, 
revealing a high density contrast, with scale length corresponding to $\sim 10^{11}$ cm, 
which is the direct result of the companion's gravitational influence in the common envelope.  In 
the immediate region surrounding the embedded companion, 
radiation pressure is comparable to gas pressure. Matter is shown to be significantly 
deflected by the companion. The velocity drops off more gradually than the density,
with a scale length $\sim 5$ times larger. In this region, typical 
velocities are of the order of 150 km s$^{-1}$, whereas the velocity of the companion is of the
order of 
$\sim 50$ km s$^{-1}$. The gas flows are subsonic, and there is no evidence for the presence 
of shocks.

We note that our calculation does not include an explicit model for the embedded boundary 
condition imposed by the surface of the companion star. A realistic inner boundary 
condition would fall somewhere between two extreme types: a fully absorbing boundary, 
corresponding to accretion onto a black hole, and a fully reflecting boundary, 
perhaps corresponding to a star with a surface that retains very little accreted 
material. In the latter case we would expect to see a clear accretion shock. Because 
we allow gas to accumulate in the companion's vicinity, 
it heats up due to compression and creates a back pressure that resists further accretion. 
Thus, our calculation is intermediate between the two cases, and the accretion rate
that we measure should be regarded as only indicative.

\begin{figure}
\epsscale{0.9}
\plotone{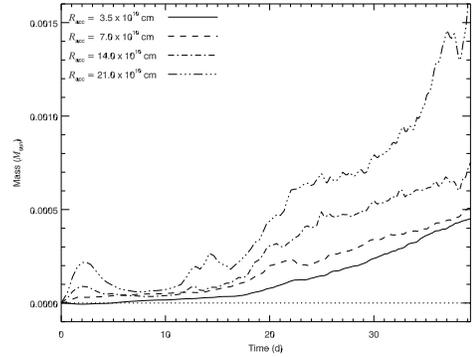}
\caption{\label{Fig:macc} Variation of mass, in units of solar mass, accumulated within a range of radii for the 
control surface centered on the embedded companion as a function of time in days.}
\end{figure}

Since the particle cloud does not provide a boundary for the gas on the 
grid, we computed the mass flux (using the respective equivalents to 
Eq.~\ref{Eqn:flux}) for a range of assumed radii.  (The accreted angular momentum 
rates could not be reliably calculated close to the companion due to insufficient 
numerical resolution. The length scale over which the pressure gradients would 
be calculated is comparable to the size of the control region.) We found that 
the mass accretion rate fluctuates significantly on timescales of less than a day. 
Thus, we integrated this rate to determine the accumulation 
of mass with time. In Figure~\ref{Fig:macc}, the accumulated mass is illustrated as a function of 
time for different control radii, measured from the companion's position, ranging from 
$3.5 \times 10^{10}$ to $2.1 \times 10^{11}$ cm.  A secular trend is evident, 
although variations are found to be larger at greater distances from the companion. The 
numerical results show that the magnitude of the accreted mass is a function of the radius 
of the control surface, with lower accreted mass associated with smaller control 
surfaces.  In particular, only about 1/3 of the matter at the outermost surface 
reached the particle cloud.  It appears that the differences in the mass accumulation 
decrease with time at smaller control surfaces, whereas they increase with time at larger 
radii, suggesting that estimates for the mass accretion rate should be based on 
evaluations at small radii.  After an evolution of 39 days,
the companion object would have accreted $\sim 5 \times 10^{-4} 
\msun$, leading to an effective mass 
accretion rate of about $0.005 \mpy$ (assuming that the angular momentum does not 
significantly impede the accretion process). During the first 35 days of evolution,
instantaneous mass accretion rates rarely
exceed $0.05 \mpy$, and then they quickly rise to $\sim 0.1 \mpy$.

Equations (\ref{Eqn:rcap}) and (\ref{Eqn:racc}) can be used to infer an expected
accretion rate $\dot{M}_{\rm BH} = \pi \racc^2 \rho \vrel$
on the basis of the gravitational capture radius formalism.
Using the companion mass for $M$ and taking $\rho$, $\vrel$, $\cs$, and $H$ as
observed in the simulation, we find $\racc \approx 8.0 \times 10^{10}$~cm and
$\dot{M}_{\rm BH} \approx 3.2 \mpy$. This estimate is significantly
larger than the instantaneous or average mass accretion rate determined from the
simulation. Although our lack of an embedded boundary condition implies that our
measured accretion rates are only indicative, we note that the rate expected in the
gravitational capture radius formalism would allow for the accretion of the
entire red giant envelope onto the companion within less than 80 days. Thus,
this formalism must significantly overestimate the real
accretion rate. The physical reason for this discrepancy is evident in
Figure~\ref{Fig:flow}. Whereas the capture radius formalism assumes a supersonic
wind flowing directly past the accretor, the common-envelope system develops a
relatively slow-moving region near the companion, and matter accretes subsonically.
Most of the gas is accelerated by the companion's potential and flung out to
large radius, where it becomes unavailable for later accretion due to its
angular momentum. In response, the companion spirals inward toward the red giant.

\section{Discussion}

We have quantitatively described the interaction of stars within a common envelope 
on the basis of the analysis of the early inspiral stage of a $1.05 \msun$ red giant 
and a $0.6 \msun$ binary companion.  The orbital decay is dominated by the nonaxisymmetric 
gravitational drag associated with the self-gravitating matter in the common envelope.  
On the basis of high-resolution three-dimensional hydrodynamic simulations, this drag is 1-2
orders of magnitude greater than the hydrodynamic drag.  As a consequence, the 
orbital decay timescale is much shorter than that derived from analyses based on the 
Hoyle-Lyttleton-Bondi picture of accretion from a uniform medium by a gravitating point 
mass moving at supersonic speeds. In this latter description the gravitational drag 
associated with an accretion wake is not significantly larger than the hydrodynamic 
drag. The effect of long-range gravitational interactions is critical for reliable estimates 
of the orbital decay timescale and energy dissipation rate.  In this picture the drag and the 
mass accretion rate are not as directly related as they are in the Hoyle-Lyttleton-Bondi-type 
description, because the dominant drag term is gravitational, rather than hydrodynamic, in 
origin.  This decoupling is reflected in the much larger ratio $L_{\rm d} / \dot{M} \sim
10^{15}-10^{16}$~cm$^2$~s$^{-2}$ measured in the simulation than expected from the
gravitational capture radius formalism ($\sim 10^{14}$~cm$^2$~s$^{-2}$).

Although the mass accretion rates estimated from our simulation should be regarded
as only indicative due to the lack of a detailed inner boundary treatment for the
companion, the structure of the flow (dominated as it is by tidal effects) strongly
suggests that the true mass accretion rate should be much smaller than the rate
expected in the gravitational capture radius formalism. The effective capture radius,
on the basis of the observed ambient density and relative velocity, is almost an order of
magnitude smaller than the expected value. In any case, the expected value would
lead to an unrealistic level of accretion over the common-envelope
period. Furthermore, the inspiral time suggested by the early evolution
of our three-dimensional simulation is much shorter than that found in earlier
one- and two-dimensional calculations that are based on the Bondi-Hoyle-Lyttleton picture. 
Assuming that the discrepancy in mass
accretion rate continues into the deep inspiral phase, the total accumulation of
matter onto the companion should be much smaller than previously expected.

These results would have little effect on the mass of an embedded main-sequence star 
because of its tendency to expand as a result of the high entropy within the common 
envelope (see Hjellming \& Taam 1991); however, it can significantly affect the outcome 
for neutron stars within a common envelope.  In particular, the 
estimated accretion rates exceed $10^{-3} \mpy$, for which steady state accretion flows 
with neutrino losses are possible (Chevalier 
1989; Houck \& Chevalier 1991). At these hypercritical mass accretion rates, photons 
are trapped in the flow and the Eddington limit is not applicable.  On the basis of this 
hypercritical accretion flow regime, Chevalier (1993) and Brown (1995) suggest that 
neutron stars embedded in the common envelope would accrete sufficient mass to form
low-mass black holes (although see Chevalier 1996), and, 
hence, the formation of binary radio pulsars would require an
evolutionary scenario involving progenitor stars of nearly equal mass (Brown 1995). 
However, the population synthesis of binary black holes and neutron stars by Belczynski 
et al.\ (2002), including hypercritical accretion, resulted in an average accretion of $0.4 
\msun$.  Such a high rate of mass accretion is inconsistent with the observed masses of binary 
radio pulsars ($\sim 1.35 \msun$; Thorsett \& Chakrabarty 1999) and indicates the need for 
a reduction in accreted matter during the common-envelope phase (see also Belczynski et al.\ 2007). Our calculations show that the necessary reduction may arise naturally as a
result of a more realistic treatment of the common-envelope phase. Consequently, this 
reduction could also lead to a reduction in the number of low-mass 
black holes, depending on the maximum mass of neutron stars, resulting from the accretion 
induced collapse of massive accreting neutron stars in the common-envelope phase.  Similarly,
the mass accretion, which was found to be as large as several solar masses for black hole
accretors, would also be reduced, thereby affecting the masses and spins of double black holes 
emerging from the common-envelope phase. 
 
Further investigations are planned to examine the generality of these results regarding 
mass accretion and to quantify the importance of these processes for determining the 
properties (mass and spin) and ultimate fate of the compact components in short-period binary 
system populations. Such studies are not only important for determining the masses of binary 
neutron star and black hole systems resulting from the common-envelope phase (Belczynski et al.\ 2007), but also their orbital periods, which directly influence the expected merger rates of 
such binary populations as sources for gravitational wave detection in the advanced LIGO 
experiment.

\acknowledgments 
P.M.R. acknowledges helpful conversations with Charles Gammie and Stu Shapiro.
This work was partially supported by the National Center for Supercomputing
Applications under allocation AST040024.
Partial support has also been provided by the NSF through grants
AST-0200876 and AST-0703950. FLASH was developed largely by the
DOE-supported ASC/Alliances Center for Astrophysical Thermonuclear Flashes at
the University of Chicago.

\end{document}